\documentclass[letterpaper]{JHEP3}

\usepackage{amsmath,amsthm, amssymb}

\newcommand{\be}{\begin{equation}}
\newcommand{\ee}{\end{equation}}
\newcommand{\ba}{\begin{array}}
\newcommand{\ea}{\end{array}}

\newcommand{\bea}{\begin{eqnarray}}
\newcommand{\eea}{\end{eqnarray}}
\newcommand{\bma}{\begin{matrix}}
\newcommand{\ema}{\end{matrix}}
\newcommand{\bpm}{\begin{pmatrix}}
\newcommand{\epm}{\end{pmatrix}}
\newcommand{\nn}{\nonumber}
\newcommand{\half}{\frac{1}{2}}
\newcommand{\mc}{\mathcal}

\newcommand{\p}{\partial}

\newcommand{\rr}{\prime}
\newcommand{\ov}{\overline}
\newcommand{\wh}{\widehat}
\newcommand{\wt}{\widetilde}

\newcommand{\psibar}{\ov \psi}

\newcommand{\labar}{\ov \la}

\newcommand{\chibar}{\ov \chi}
\newcommand{\tabar}{\ov \ta}

\newcommand{\ep}{\epsilon}

\newcommand{\al}{\alpha}

\newcommand{\la}{\lambda}
\newcommand{\da}{\delta}

\newcommand{\Ga}{\Gamma}
\newcommand{\ga}{\gamma}

\newcommand{\si}{\sigma}
\newcommand{\ta}{\theta}

\newcommand{\qrq}{\quad\Rightarrow\quad}

\newcommand{\epbar}{\ov\ep}

\title{Rigid supersymmetry with boundaries}
\author{ Dmitry V.~Belyaev \\
Deutsches Elektronen-Synchrotron, DESY-Theory\\ 
Notkestrasse 85, 22603 Hamburg, Germany\\
E-mail: \email{dmitry.belyaev@desy.de}
}
\author{ Peter van Nieuwenhuizen\\
C.~N.~Yang Institute for Theoretical Physics, SUNY at Stony Brook \\
Stony Brook, NY 11794-3840, USA\\
E-mail: \email{vannieu@max2.physics.sunysb.edu}
}
\date{\today}

\preprint{DESY 08-\$\$\$ \\ YITP-SB-08-01}

\abstract{
We construct rigidly supersymmetric bulk-plus-boundary actions,
both in $x$-space and in superspace. 
For each standard supersymmetric
bulk action a minimal supersymmetric bulk-plus-boundary action
follows from an extended $F$- or $D$-term formula. Additional
separately supersymmetric boundary actions can be systematically
constructed using co-dimension one multiplets 
(boundary superfields).
We also discuss the orbit of boundary conditions which follow
from the Euler-Lagrange variational principle.
}

\begin{document}

\numberwithin{equation}{section}


\newpage
\section{Introduction} 

Since its beginning, research in supersymmetry (susy) has mainly
been concerned with constructing invariant actions, and deducing
the consequences of their field equations. However, the field
equations are only half of the information one needs for a 
mathematically well-posed problem; the other half are the boundary
conditions (BC) one must impose on the fields.
In susy (and supergravity) one usually assumes that fields
fall off sufficiently fast at (spacelike and timelike) infinity
and that boundary terms which arise from partial integration
may be omitted. However, if there is a boundary, this assumption
is unwarranted, and one must face the issue of BC.
In this article we present a thorough study of BC in models
of rigid susy with a timelike boundary.

We distinguish between two kinds of BC: those which are needed to keep
the action invariant under rigid susy, and those which arise from
the Euler-Lagrange (EL) field equations.
The first set is off-shell, the second set is on-shell.
Our main philosophy is to construct bulk-plus-boundary
actions which are susy by themselves (under half of bulk susy),
so no BC are needed to cancel boundary terms in the
susy variation of the action. (This approach was first
advocated in \cite{db1,db2}.)
We call such models ``susy without BC.''
We develop an extension of the usual tensor calculus which gives
the boundary action which one must add to the bulk action to
obtain ``susy without BC.'' 
Once this boundary action has been constructed, 
one can study the EL variation
of the bulk-plus-boundary action. In the bulk it gives
standard field equations, but boundary terms arise which can only
be canceled by imposing BC on some of the fields.
It follows that the BC one obtains in this way are, to begin with,
BC on on-shell fields. However, once a set of such BC has been obtained,
one can also require that they hold for off-shell fields.
For example, in a path integral approach where fields are, of course,
off-shell, we might still impose such BC on these off-shell fields.
We shall first study the various possibilities
in the examples considered below, and come back to more definite
statements in the conclusions.

As always, one has the option of using the $x$-space (component)
approach, or the superspace approach. In an earlier article \cite{dbpvn1}
we analyzed a particular supergravity model ($N=1$ supergravity
in $2+1$ dimensions), and since the superspace approach for
supergravity is rather complicated, we cast that article entirely
in $x$-space. However, the superspace approach of rigid susy is
much simpler, and thus we shall first derive our new results in
$x$-space, but then recast these results into superspace.

Our program of constructing invariant actions consists of two parts.
First we obtain actions with ``susy without BC'' by adding suitable
actions on the boundary; these boundary actions are not susy by themselves
but merely complete the bulk actions, and we shall have to find
an appropriate superspace description for them. 
Next, for some models it will turn out that we
need to construct another action on the boundary
which is susy by itself; this action can be described by $x$-space
or superspace methods in one dimension less (boundary superfields).

Before introducing our extension of the tensor calculus, it may
be helpful to point out some possible pitfalls.
First, the boundary terms one obtains from partially integrating
terms in the susy variation of the action are in general different
from those in the EL variation of the actions.
Thus even if ``susy without BC'' holds, one will in general nead EL BC.
Second, BC on spacelike surfaces (initial conditions) have physically
a very different meaning from BC on timelike surfaces
(genuine BC, at all times).
We consider only the latter,
and choose as boundary the hypersurface at $x^3=0$.
However, from a space-time point of view, one can treat these
two sets of BC on equal footing; 
technically this is achieved by introducing projection operators
$P_\pm=\half(1\pm n^\mu\ga_\mu)$ where $n^\mu$ is the normal to
the boundary, and decomposing the susy parameters into eigenspinors
$\ep_\pm$ of this projection operator.
This procedure was used in \cite{VV}, but note that in that article
a very different philosophy was used: 
no ``susy without BC'' was implemented,
and the consistency of the complete set of susy BC
and EL BC was studied (the ``orbit of BC'').
Since (half of) bulk susy is unbroken in our case, 
and auxiliary fields are present,
the study of the
orbit of BC can be written as BC on boundary superfields.
Finally, it is of course true that in varying actions on the boundary
one may again need to partially integrate, thus obtaining boundary
terms on the boundary. We assume that all total derivatives on the
boundary vanish. This is not necessary, but it simplifies the analysis.

Let us now introduce our extension of the usual tensor calculus
which takes boundaries into account. As an example, consider the
usual $F$-term formula for an invariant action in the bulk.
Decomposing the integration measure $d^{m+1}x$ into a measure $d^m x$
on the boundary and $dx^3$ away from the boundary, one has
\bea
S=\int_\mc{M} dx^3 d^m x F
\eea
Since $F$ varies into a total derivative, $\da F=\epbar\ga^\mu\p_\mu\psi$,
the variation of $S$ is equal to a boundary term 
$\da S=-\int d^m x (\epbar\ga^3\psi)$.
We shall introduce a susy parameter $\ep_{+}$ satisfying 
$\epbar_{+}\ga^3=-\epbar_{+}$. 
Then $\da S=\int d^m x \, \epbar_{+}\psi$, and since $\da A=\epbar\psi$,
we find a suitable action $S_\text{boundary}=\int d^m x \, A$
on the boundary, whose $\ep_{+}$ 
variation cancels the variation of the bulk
action. So, the usual $F$-term formula is extended to the
following ``$F+A$'' formula for a bulk-plus-boundary 
action\footnote{
In \cite{dbpvn1} we derived the analog of this ``$F+A$''
formula in supergravity. There, instead of \emph{components} 
$F$~and~$A$,
one needs to use the corresponding \emph{densities}.
}
\bea
\label{introFA}
S=\int_\mc{M} dx^3 d^m x \, F -\int_{\p\mc{M}} d^m x \, A
\eea 
We will find that this extended $F$-term formula works both in
3 and 4 dimensions. In what follows, we will indicate other
ways in which this formula can be derived and will apply it to
various models of rigid susy in 3 and 4 dimensions.

The presence of boundary terms may modify EL BC.
Consider as an example the action for the spinning string, which
is the same as the 2D Wess-Zumino (WZ) action,
\bea
S_\text{WZ}=\int d\si dt \mc{L}, \quad
\mc{L}=-\p_\mu X\p^\mu X-\psibar\ga^\mu\p_\mu\psi +F^2
\eea
The susy variation
$\da\mc{L}=\p_\mu(-\epbar\ga^\mu\ga^\nu\psi\p_\nu X
+\epbar\ga^\mu\psi F)$
leads to a boundary term at $\si=0$
which can be canceled (when $\ep=\ep_{+}$)
by adding the following boundary action at $\si=0$,
\bea
S_b=-\int dt \Big[ X F+X \p_\si X \Big]
\eea
(The first term is the term denoted by ``$A$'' in (\ref{introFA})
while the second term is produced if one rewrites the action
$X\p_\mu\p^\mu X$ as obtained from the tensor calculus as
$-\p_\mu X\p^\mu X+\p_\si(X\p_\si X)$ and uses
$\int d\si \p_\si(X\p_\si X)=-X\p_\si X$.)
From the EL variation of $S_{WZ}$, 
if one requires that all coefficients of 
varied fields vanish, one obtains a set of EL BC which is
too strong,
\bea
X=F-\p_\si X=\psi_{+}=\psi_{-}=0
\eea

As explained below (see section \ref{sec-3EL}) for the 3D WZ model,
which is very similar to the 2D WZ model, one can add a separately
susy action on the boundary,
\bea
S_\text{b}(extra)=\int dt \Big(
X F+ X\p_\si X-\half\psibar\psi \Big)
\eea
The total string action now becomes
\bea
S=S_\text{WZ}-\half\int dt \psibar\psi
\eea
and one finds now the same EL BC for X as before, $\da X\p_\si X=0$,
while for $\psi$ one finds $\psibar_{+}\da\psi_{-}=0$.
These are the usual Dirichlet or Neumann conditions for $X$ and the 
Neveu-Schwarz or Ramond conditions for $\psi$.\footnote{
For the 2D case, our conventions give $x^3=\si$ and $\ga^3=\ga^\si$.
Taking a particular representation of gamma matrices (which we
avoid in this paper) one can rewrite our two-component spinors
$\psi_{\pm}$ in terms of one-component spinors $\psi^{\pm}$ 
and recover the usual form of the NS and R conditions,
$\psi^{+}=\pm\psi^{-}$ (see e.g. \cite{LRN}).
}
They are needed to make the EL variation of $S$ vanish on-shell,
but they are not needed to make the $\ep_{+}$ susy variation of
$S$ vanish (off-shell).

The EL variations on the boundary are of the form ``$p \da q$,''
and one might expect that one might choose 
either $p=0$ or $q=\text{const}$
as BC for each field (which would give $2^N$ sets of BC where
$N$ is the number of $q$'s).
However, this is incorrect: consistency of the EL BC with susy
\cite{pdv,igarashi,LRN,VV} leaves only two families of BC
\cite{hawking,VV}.
These families become shorter when auxiliary fields are
properly incorporated. Then, as we show in section \ref{sec-3EL},
each family corresponds to a BC on a boundary superfield
\cite{db1,db3}. This nice result is of course due to
``susy without BC.''

We remark that
our ``susy without BC'' approach, and the ``$F+A$'' formula 
in particular, can be applied to a variety of physically interesting
models, including those involving strings and branes, solitons
and instantons. Some of the applications were discussed in 
\cite{dbpvn1}.

\newpage
\section{Extended tensor calculus} 

In this section, we present extensions of the standard $F$-
and $D$-term formulae for the 3D and 4D cases.\footnote{
Our spinors are Majorana spinors, so $\psibar\equiv\psi^\dagger i\ga^0$
is equal to $\psibar=\psi^T C$ where $C\ga^\mu C^{-1}=-(\ga^\mu)^T$
and $C^T=-C$. Furthermore, $\ga^\mu\ga^\nu=\eta^{\mu\nu}+\ga^{\mu\nu}$
with $\eta^{\mu\nu}=(-1,+1,\dots,+1)$ and in $d=4$ we use
$\ga_5$ with $\ga_5^2=1$.
}
In both dimensions, we use Cartesian coordinates $x^\mu$
to describe the bulk $\mc{M}$ and assume that the boundary $\p\mc{M}$
is at $x^3=0$ and is parametrized by $x^m$. In $\mc{M}$, $x^3>0$.
In the presence of the boundary, half of susy is (spontaneously)
broken. We choose to preserve the half parametrized by
$\ep_{+}=P_{+}\ep$, where $P_{\pm}=\half(1\pm\ga^3)$.
Then $\ga^3\ep_{+}=\ep_{+}$ but $\epbar_{+}\ga^3=-\epbar_{+}$.

\subsection{3D extended $F$-term formula} 

Consider the 3D $N=1$ scalar multiplet $\Phi_3=(A,\psi,F)$,
\bea
\label{tr3sc}
\da A=\epbar\psi, \quad
\da\psi=\ga^\mu\ep\p_\mu A+F\ep, \quad
\da F=\epbar\ga^\mu\p_\mu\psi
\eea
The standard $F$-term formula gives a bulk action 
$\int_\mc{M} d^3x F$ that is not susy in the presence of
the boundary. Its susy variation gives rise to a boundary term
$-\int_{\p\mc{M}} d^2x (\epbar\ga^3\psi)$.
Our extended $F$-term formula,
\bea
\label{F3}
S=\int_\mc{M} d^3x F-\int_{\p\mc{M}} d^2x A
\eea
gives a bulk-plus-boundary action that is invariant under 
$\ep_{+}$ susy. Indeed,
$\da F=\p_\mu(\epbar\ga^\mu\psi)$ yields a boundary term
$-\epbar_{+}\ga^3\psi=\epbar_{+}\psi$ under
the $\ep_{+}$ susy.
Clearly, the corresponding variation of $A$ on the boundary,
$\da A=\epbar_{+}\psi$, cancels the contribution
from the bulk.

Another 3D $N=1$ multiplet, which includes a 3D vector $v_\mu$, 
is the spinor multiplet $\Psi_3=(\chi,M,v_\mu,\la)$
with the following susy transformation rules,
\bea
\label{tr3sp}
&& \da\chi=M\ep+\ga^\mu\ep v_\mu, \quad
\da M=-\half\epbar\la+\epbar\ga^\mu\p_\mu\chi \nn\\
&& \da v_\mu=-\half\epbar\ga_\mu\la+\epbar\p_\mu\chi, \quad
\da\la=2\ga^{\mu\nu}\ep\p_\mu v_\nu
\eea
The highest component, $\la$, transforms into a total derivative
but, as $\la$ is a fermion, we cannot use a ``$\la$-term formula''
for constructing susy actions.

\subsection{4D extended $F$- and $D$-term formulae} 

For the 4D $N=1$ scalar (chiral) multiplet
$\Phi_4=(A,B,\psi,F,G)$,
\bea
&& \da A=\epbar\psi, \quad
\da B=-i\epbar\ga_5\psi \nn\\
&& \da\psi=\ga^\mu\p_\mu(A-i\ga_5 B)\ep+(F+i\ga_5 G)\ep \nn\\
&& \da F=\epbar\ga^\mu\p_\mu\psi, \quad
\da G=i\epbar\ga_5\ga^\mu\p_\mu\psi
\eea
the extended $F$-term formula is
\bea
\label{F4}
S=\int_\mc{M} d^4x F-\int_{\p\mc{M}} d^3x A
\eea
Alternatively, one can use the extended $G$-term formula,
\bea
S=\int_\mc{M} d^4x G-\int_{\p\mc{M}} d^3x B
\eea
In both cases, we find a bulk-plus-boundary action that is
invariant under $\ep_{+}$ susy.

For the 4D $N=1$ vector multiplet
$V_4=(C,\chi,H,K,v_\mu,\la,D)$, 
\bea
&& \da C=i\epbar\ga_5\chi, \quad
\da\chi=(i\ga_5 H-K-\ga^\mu v_\mu+i\ga^\mu\ga_5\p_\mu C)\ep \nn\\
&& \da H=i\epbar\ga_5\ga^\mu\p_\mu\chi+i\epbar\ga_5\la, \quad
\da K=-\epbar\ga^\mu\p_\mu\chi-\epbar\la \nn\\
&& \da v_\mu=-\epbar\p_\mu\chi-\epbar\ga_\mu\la, \quad
\da\la=\ga^{\mu\nu}\ep\p_\mu v_\nu+i\ga_5 D\ep, \quad
\da D=i\epbar\ga_5\ga^\mu\p_\mu\la
\eea
the highest component, $D$, transforms into a total derivative
and the standard $D$-term formula gives a bulk action
$S=\int_\mc{M} d^4x D $. This action is not susy in the
presence of the boundary. 
Our extended $D$-term formula is
\bea
\label{D4}
S=\int_{\mc{M}} d^4x D  + \int_{\p\mc{M}} d^3x (H-\p_3 C) 
\eea
and it gives bulk-plus-boundary actions that are invariant
under $\ep_{+}$ susy. (Here and hereafter we assume that total 
tangential $\p_m$ derivatives integrated over the boundary vanish.)

The extended $D$-term formula can be derived from the extended
$F$-term formula. Indeed, given a vector multiplet $V_4$, we
can construct the following scalar multiplet,
\bea
\Phi_4[V_4]=(-H, \; K, \; -i\ga_5(\la+\ga^\mu\p_\mu\chi), \;
D+\p^\mu\p_\mu C, \; -\p^\mu v_\mu) 
\eea
Applying (\ref{F4}) to this multiplet, we recover (\ref{D4}).
Clearly, the $F$-term formula covers all cases, and is also
simpler.

\newpage
\section{Applications} 

The extended $F$- and $D$-term formulae of the previous section
can be applied to a variety of composite multiplets. 
This allows straightforward construction of susy bulk-plus-boundary 
actions that are minimal extensions of known (bulk) actions.
In this section, we will consider several examples of this procedure.
Generically, terms linear in the (bulk) auxiliary 
fields appear in the boundary actions. We will find that in some, 
but not all, cases these terms can be eliminated by adding 
separately susy boundary actions. It will follow that,
generically, ``susy without BC'' requires the presence of auxiliary
fields.

\subsection{3D Wess-Zumino model} 

Given a 3D scalar multiplet $\Phi_3(A)=(A,\psi,F)$, we can
construct a ``kinetic'' scalar multiplet whose lowest component
is $F$,
\bea
\label{Tphi3}
T(\Phi_3)\equiv\Phi_3(F)=(F, \; \ga^\mu\p_\mu\psi, \; \p_\mu\p^\mu A)
\eea
The product of $\Phi_3(A)$ and $\Phi_3(F)$
gives another 3D scalar multiplet,
\bea
\Phi_3(A F)=(A F, \; F\psi+A\ga^\mu\p_\mu\psi, \;
F^2+A\p_\mu\p^\mu A-\psibar\ga^\mu\p_\mu\psi)
\eea
Applying our extended $F$-term formula (\ref{F3}) to this multiplet,
we find the following bulk-plus-boundary action,
\bea
\label{3DWZ}
S=\int_\mc{M} d^3x (F^2-\p_\mu A\p^\mu A-\psibar\ga^\mu\p_\mu\psi)
-\int_{\p\mc{M}} d^2x (A F+A\p_3 A)
\eea
where we partially integrated to arrive at the standard form for
the bulk action. This is the 3D Wess-Zumino (WZ) model supplemented
by a particular boundary term. The bulk-plus-boundary action is,
by construction, invariant under $\ep_{+}$ susy, as can be explicitly
verified.

We observe that the bulk auxiliary field $F$ appears linearly
on the boundary. Therefore, eliminating $F$ via its field
equation would require imposing a boundary condition $A=0$.
The action without auxiliary fields 
would then necessarily be ``susy with BC.''
To be able to eliminate $F$ while preserving ``susy without BC,'' 
we will now look for a separately susy boundary action that cancels the term 
linear in $F$. 

Separately susy boundary actions can be constructed systematically
using co-dimension one multiplets. To this extent, we split the
3D $N=1$ multiplet $\Phi_3$ into two 2D $N=(1,0)$ multiplets 
under the $\ep_{+}$ susy.
(The third coordinate, $x^3$, will appear in the 2D multiplets 
as a parameter.) Defining $\psi_{\pm}=P_{\pm}\psi$, we find
\bea
\label{tr2sf}
&& \da A=\epbar_{+}\psi_{-}, \quad
\da\psi_{-}=\ga^m\ep_{+}\p_m A \nn\\
&& \da\psi_{+}=(F+\p_3 A)\ep_{+}, \quad
\da(F+\p_3 A)=\epbar_{+}\ga^m\p_m\psi_{+}
\eea
so that we find a scalar and a spinor 2D $N=(1,0)$ 
multiplet,\footnote{
The highest component of $\Psi_2$, $F+\p_3 A$, transforms
into a total $\p_m$ derivative under $\ep_{+}$ susy.
Integrating this component over the 3D manifold $\mc{M}$ with
boundary $\p\mc{M}$ gives our ``$F+A$'' formula (\ref{F3}).
The four-dimensional extended $F$- and $D$-term formulae can
also be derived in such a way using co-dimension one multiplets.
}
\bea
\label{cm1}
\Phi_2=(A, \; \psi_{-}), \quad
\Psi_2=(\psi_{+}, \; F+\p_3 A)
\eea
Their product is another spinor multiplet,
\bea
\Phi_2 \times \Psi_2 =(A\psi_{+}, \;\;
A(F+\p_3 A)-\psibar_{+}\psi_{-})
\eea
whose highest component transforms into a total $\p_m$ derivative.
Therefore, the following action
\bea
\label{delWZ3}
\int_{\p\mc{M}} d^2x (A F+A\p_3 A-\psibar_{+}\psi_{-})
\eea
is invariant under $\ep_{+}$ susy. 
Adding it to (\ref{3DWZ}), the first two terms cancel, 
and we obtain
\bea
\label{3DWZa}
S=\int_\mc{M} d^3x (F^2-\p_\mu A\p^\mu A-\psibar\ga^\mu\p_\mu\psi)
-\int_{\p\mc{M}} d^2x \half\psibar\psi
\eea
where we used $\psibar\psi=2\psibar_{+}\psi_{-}$.
Setting $F=0$ in the action and susy transformations, we arrive
at the 3D WZ model without auxiliary fields that is still
``susy without BC.''

\subsection{4D Wess-Zumino model} 

The 4D WZ model will turn out to be more subtle.
We start again with the scalar multiplet
$\Phi_4=(A,B,\psi,F,G)$ and construct the kinetic multiplet,
\bea
T(\Phi_4)=(F, \; -G,\; \ga^\mu\p_\mu\psi, \; 
\Box A, \; -\Box B)
\eea
where $\Box=\p_\mu\p^\mu$. Their product gives the following
scalar multiplet
\bea
\Phi_4\times T(\Phi_4) &=& (A F+B G, \quad -A G+B F, \quad 
(A+i\ga_5 B)\ga^\mu\p_\mu\psi+(F-i\ga_5 G)\psi, \nn\\
&& \hspace*{-30pt}
A\Box A+B\Box B+F^2+G^2-\psibar\ga^\mu\p_\mu\psi, \quad
-A\Box B+B\Box A+i\psibar\ga_5\ga^\mu\p_\mu\psi)
\eea
Applying our extended $F$-term formula (\ref{F4}) to this
multiplet, we find, after some partial integration, the following
action
\bea
\label{4DWZ}
S &=& \int_\mc{M} d^4x \Big[
-\p_\mu A \p^\mu A-\p_\mu B\p^\mu B
-\psibar\ga^\mu\p_\mu\psi+F^2+G^2 \Big]
\nn\\
&& -\int_{\p\mc{M}} d^3x \Big[A(F+\p_3 A)+B(G+\p_3 B)\Big]
\eea
This action is invariant under $\ep_{+}$ susy by construction.
As in the 3D case, we find terms linear in the bulk auxiliary
fields, $F$ and $G$, in the boundary action.
However, unlike the 3D case, we will find that one cannot eliminate 
both terms by adding a separately susy boundary action.

To construct separately susy boundary actions,
we split the 4D $N=1$ scalar multiplet $\Phi_4$ into two 3D $N=1$ 
scalar multiplets under $\ep_{+}$ susy. (The fourth coordinate, $x^3$,
will appear in the 3D multiplets as a parameter.)
Defining $\psi_{\pm}=P_{\pm}\psi$, we find
\bea
& \da A=\epbar_{+}\psi_{-}, \quad
\da\psi_{-}=\ga^m\ep_{+}\p_m A
+i\ga_5(G+\p_3 B)\ep_{+}, \quad
\da(G+\p_3 B)=i\epbar_{+}\ga_5\ga^m\p_m\psi_{-} &
\nn\\
& \da B=-i\epbar_{+}\ga_5\psi_{+}, \quad
\da\psi_{+}=i\ga_5\ga^m\ep_{+}\p_m B
+(F+\p_3 A)\ep_{+}, \quad
\da(F+\p_3 A)=\epbar_{+}\ga^m\p_m\psi_{+} &
\nn
\eea
so that the two 3D multiplets contained in $\Phi_4$ are\footnote{
We keep here the 4D gamma matrices to describe 3D multiplets.
This description avoids explicit decomposition of the gamma
matrices at the price of an unusual definition of susy transformations.
}
\bea
\label{cm2}
\Phi_A=(A, \; \psi_{-}, \; G+\p_3 B), \quad
\Phi_B=(B, \; -i\ga_5\psi_{+}, \; -F-\p_3 A)
\eea
Their product yields another 3D scalar multiplet
\bea
\Phi_A \times \Phi_B = \Big(A B, \quad
-i\ga_5 A\psi_{+}+B\psi_{-}, \quad
-A (F+\p_3 A)+B(G+\p_3 B)+\psibar_{+}\psi_{-} \Big)
\eea
The highest component of this multiplet can be used to
construct the following separately susy boundary actions,
\bea
\label{del4DWZ}
\al\int_{\p\mc{M}} d^3x \Big[
-A(F+\p_3 A)+B(G+\p_3 B)+\half\psibar\psi \Big]
\eea
where we used $\psibar\psi=2\psibar_{+}\psi_{-}$.
We observe that adding this action to (\ref{4DWZ}) with 
$\al=-1$ or $\al=+1$, we can cancel
either the term linear in $F$ or the term linear in $G$,
but not both.
Therefore, eliminating auxiliary fields in the 4D WZ model
cannot be done while maintaining the ``susy without BC''
property (as a boundary condition $A=0$ or $B=0$ arises in
the process). Turning this around, we see that generically
auxiliary fields are required for ``susy without BC.''

\subsection{3D Maxwell model} 

For the 3D $N=1$ spinor multiplet $\Psi_3=(\chi,M,v_\mu,\la)$,
in (\ref{tr3sp}),
the kinetic multiplet $T(\Psi_3)$ is a spinor multiplet whose
lowest component is $\la$,
\bea
\label{Tpsi3}
T(\Psi_3)\equiv\Psi_3(\la)=(\la, \;\; 0, \;\;
-\ep_{\mu\nu\rho}F^{\nu\rho}, \;\; 2\ga^\mu\p_\mu\la)
\eea
where $F_{\mu\nu}=\p_\mu v_\nu-\p_\nu v_\mu$ and 
$\ga^{\mu\nu\rho}=-\ep^{\mu\nu\rho}$
(then $-\ga^\mu\ep_{\mu\nu\rho}=\ga_{\nu\rho}$).
We will take $\ep^{013}=+1$ so that 
\bea
\label{ga013}
\ga^0\ga^3=\ga^1=\ga_1, \quad
\ga^1\ga^3=\ga^0=-\ga_0
\eea
The product of $T(\Psi_3)$ with itself gives the following
3D $N=1$ scalar multiplet,
\bea
\label{3DLL}
T(\Psi_3)\times T(\Psi_3)
=(\labar\la, \;\; -2\ga^{\mu\nu}\la F_{\mu\nu}, \;\;
4 F_{\mu\nu}F^{\mu\nu}+2\labar\ga^\mu\p_\mu\la)
\eea
Applying our ``$F+A$'' formula (\ref{F3}) to this
multiplet, we obtain
\bea
S=\int_{\mc{M}} d^3x \Big[
4 F_{\mu\nu}F^{\mu\nu}+2\labar\ga^\mu\p_\mu\la \Big] 
-\int_{\p\mc{M}} d^2x (\labar\la)
\eea
This action is invariant under $\ep_{+}$ susy and
contains the usual susy Maxwell action in the bulk
(up to an overall normalization constant).\footnote{
One would have to define $\la^\text{new}=\la/2$ to get
canonical kinetic terms for $v_\mu$ and $\la^\text{new}$ 
at the same time.
Our choice of $\la$ in 3D followed from a natural parametrization of
the corresponding superfield $\Ga_\al$, see (\ref{phiga}).
In 4D, our $\la$ is canonically defined.
} 
As there is no auxiliary field in this action, there is no
particular reason to add a separately susy boundary action.

\subsection{4D Maxwell model} 

The Maxwell action for the 4D $N=1$ vector multiplet 
$V_4=(C, \chi, H, K, v_\mu, \la, D)$
can be written using the $F$-term formula applied to the following
composite scalar multiplet,
\bea
\label{4DLL}
\Phi_4 &=& (A_c, \; B_c, \; \psi_c, \; F_c, \; G_c)
\\
&=& (\labar\la, \;\; \labar i\ga^5\la, \;\;
-\ga^{\mu\nu}F_{\mu\nu}\la+2i\ga^5\la D, \;\;
F_{\mu\nu}F^{\mu\nu}+2\labar\ga^\mu\p_\mu\la -2 D^2, \;\;
\half\ep^{\mu\nu\rho\si}F_{\mu\nu}F_{\rho\si}) \nn
\eea
Our extended $F$-term formula (\ref{F4}) applied to this 
multiplet gives (up to our factor $-1/2$)
\bea
\label{4Ma}
S=\int_{\mc{M}} d^4x \Big[
-\half F_{\mu\nu}F^{\mu\nu}-\labar\ga^\mu\p_\mu\la+D^2 \Big] 
+\int_{\p\mc{M}} d^3x \half\labar\la
\eea
This bulk-plus-boundary action is ``susy without BC'' by
construction.
The auxiliary field $D$ appears only in the bulk and
we can eliminate it by its field equation (set $D=0$) while
preserving the ``susy without BC'' property.
Adding a separately susy boundary action in this case is, therefore,
not required.

It is instructive, however, to discuss an alternative derivation
of the same action. As in the 3D case, we can first construct the
kinetic multiplet $T(V_4)$, a composite 4D vector multiplet 
whose lowest component is $D$,
\bea
T(V_4)=(D, \; \ga^\mu\p_\mu\la, \; 0, \; 0, \;
-\p^\mu F_{\mu\nu}, \; -\p_\mu\p^\mu\la, \; -\p_\mu\p^\mu D)
\eea
Unlike the 3D case, however, the 4D Maxwell action arises not
from $T(V_4)\times T(V_4)$, but from $V_4 \times T(V_4)$.
The latter is a composite 4D vector multiplet whose lowest
component is $\wt C=C D$. Among other components of 
$\wt V_4=V_4 \times T(V_4)$ we find, in particular,
\bea
\wt H &=& H D-\half\chibar\ga^\mu\p_\mu\la
\nn\\
\wt D &=& D^2-\half F_{\mu\nu}F^{\mu\nu}-\labar\ga^\mu\p_\mu\la
+\p_\mu\Big[ -C\p^\mu D+\half\chibar\ga^\mu\ga^\nu\p_\nu\la
+F^{\mu\nu}v_\nu \Big]
\eea
Our extended $D$-term formula (\ref{D4}) applied to $\wt V_4$
gives
\bea
\label{4Mb}
\wt S &=& \int_\mc{M} d^4x \Big[
-\half F_{\mu\nu}F^{\mu\nu}-\labar\ga^\mu\p_\mu\la+D^2 \Big] 
\nn\\&&
-\int_{\p\mc{M}} d^3x \Big[ 
v^m F_{3m}-D(H-\p_3 C)-\chibar_{+}\ga^\mu\p_\mu\la \Big]
\eea
This bulk-plus-boundary action is ``susy without BC'' by
construction. We see, however, that the boundary action contains
now a term linear in the auxiliary field $D$. Again, we would like
to find a separately susy boundary action which, upon adding it to
$\wt S$, would cancel this term. A systematic search for such an
action would require decomposing $V_4$ into co-dimension one (3D)
multiplets and then using tensor calculus to construct a 3D scalar
multiplet whose $F$ component contains the $D(H-\p_3 C)$ combination.
Instead of following this tedious procedure, we simply deduce the
answer by noting that since both $S$ and $\wt S$, in (\ref{4Ma})
and (\ref{4Mb}), are ``susy without BC,'' so is their difference,
\bea
S-\wt S=\int_{\p\mc{M}} d^3x \Big[ \half\labar\la
+v^m F_{3m}-D(H-\p_3 C)-\chibar_{+}\ga^\mu\p_\mu\la \Big]
\eea
One can verify that this boundary action is, indeed, invariant
under $\ep_{+}$ susy.

\subsection{3D Chern-Simons model} 

Returning to the 3D case and
taking now the product of $\Psi_3$ with $T(\Psi_3)$, we find
another 3D $N=1$ scalar multiplet
\bea
\Psi_3 \times T(\Psi_3)=\Big(\chibar\la, \quad
(M-\ga^\mu v_\mu)\la-\ga^{\mu\nu}F_{\mu\nu}\chi, \quad
\p_\mu(\chibar\ga^\mu\la)+2\ep^{\mu\nu\rho}v_\mu F_{\nu\rho}
+\labar\la \Big)
\eea
The 3D extended $F$-term formula (\ref{F3}) now gives
\bea
S=\int_{\mc{M}} d^3x \Big[
\p_\mu(\chibar\ga^\mu\la)+2\ep^{\mu\nu\rho}v_\mu F_{\nu\rho}
+\labar\la \Big] 
-\int_{\p\mc{M}} d^2x (\chibar\la) 
\eea
Using $\chibar\la=\chibar_{+}\la_{-}+\chibar_{-}\la_{+}$
and $\chibar\ga^3\la=-\chibar_{+}\la_{-}+\chibar_{-}\la_{+}$,
the action simplifies to
\bea
\label{CS}
S=\int_{\mc{M}} d^3x \Big[
2\ep^{\mu\nu\rho}v_\mu F_{\nu\rho}+\labar\la \Big] 
-\int_{\p\mc{M}} d^2x (2\chibar_{-}\la_{+}) 
\eea
By construction, this action is invariant under $\ep_{+}$ susy.
Its bulk Lagrangian $2\ep^{\mu\nu\rho}v_\mu F_{\nu\rho}+\labar\la$
is the usual 3D susy Chern-Simons Lagrangian.\footnote{
Combining the Maxwell and Chern-Simons actions
would require introducing a dimensionful
(mass) parameter. In fact, the Chern-Simons action gives rise
naturally to a gauge-invariant (up to a boundary variation)
mass term for the 3D vector field $v_\mu$.
}
In this model (unlike the Maxwell case)
$\la$ is nonpropagating.
We observe that it appears quadratically in the bulk and 
linearly on the boundary.
On the other hand, $\chi$ and $M$ are the fields that would be
set to zero in the 3D Wess-Zumino (WZ) gauge.\footnote{
The usual gauge transformation, $\da_g v_\mu=\p_\mu A$, is extended
in superspace to $\da\Psi_3=D\Phi_3$, see (\ref{gtr}), 
where $\Phi_3=(A,\psi,F)$ is now a multiplet of parameters. 
This gives $\da\chi=\psi$, $\da M=F$, $\da v_\mu=\p_\mu A$ and $\da\la=0$.
If this transformation is a symmetry of an action, one can impose
the WZ gauge: set $\chi=M=0$. However, (\ref{CS}) is not invariant
under such transformation (though its variation is only a boundary
term).
}
As $\chi$ appears in
the boundary action, we conclude that, generically, 
``susy without BC'' requires such fields to be present. 
In other words, it may not be possible to impose the WZ gauge and
to still have a bulk-plus-boundary action that is ``susy without BC.''

It is instructive to consider two cases when the boundary
action in (\ref{CS}) vanishes. The first case is when the WZ gauge 
$\chi=M=0$ is imposed. The bulk action then varies into\footnote{
As is well-known, in the WZ gauge one must add a compensating
gauge transformation to the susy transformation.
To keep $\da\chi=\da M=0$, we need a gauge transformation with
$(A_c,\psi_c,F_c)=(0,-\ga^\mu\ep v_\mu, \half\epbar\la)$.
The resulting susy transformations, as follows from (\ref{tr3sp}), 
are
$\da v_\mu=-\half\epbar\ga_\mu\la$ and 
$\da\la=\ga^{\mu\nu}F_{\mu\nu}\ep$.
}
\bea
\da S=\int_{\p\mc{M}} d^2x \Big[
-2(\epbar_{+}\ga^m\la_{+})v_m \Big]
\eea
The second case is when one eliminates the auxiliary field $\la$
by its field equation (that is, by setting $\la=0$
in (\ref{tr3sp}) and (\ref{CS})). 
The bulk action now varies into
\bea
\da S=\int_{\p\mc{M}} d^2x \Big[
-2(\epbar_{+}\ga^{mn}\chi_{-})F_{mn} \Big]
\eea
We see that in both cases one needs to impose some boundary conditions
to make the susy variation vanish, that is only ``susy with BC''
is possible when some of the ``auxiliary'' fields are absent.

Returning to the action (\ref{CS}), we now ask if it is possible
to find a separately susy boundary action that allows to remove
the term linear in the auxiliary field $\la$. Once again, to
construct such an action we split the 3D $N=1$ spinor multiplet 
$\Psi_3$ into 2D $N=(1,0)$ multiplets under $\ep_{+}$ susy.
First, we find that
\bea
&& \da\chi_{+}=\ep_{+}(M+v_3), \quad
\da(M+v_3)=\epbar_{+}\ga^m\p_m\chi_{+}
\nn\\
&& \da v_3=\epbar_{+}(\half\la_{-}+\p_3\chi_{-}), \quad
\da(\half\la_{-}+\p_3\chi_{-})=\ga^m\ep_{+}\p_m v_3
\nn\\
&& \da\chi_{-}=\ga^m\ep_{+}v_m, \quad
\da v_m=-\half\epbar_{+}\ga_m\la_{+}+\epbar_{+}\p_m\chi_{-}, \quad
\da\la_{+}=\ga^{mn}F_{mn}\ep_{+}
\eea
which gives a spinor multiplet $(\chi_{+},\; M+v_3)$,
a scalar multiplet $(v_3, \; \half\la_{-}+\p_3\chi_{-})$
and a multiplet $(\chi_{-}, \; v_m, \; \half\la_{+})$. 
The latter is, in fact, further reducible under $\ep_{+}$ susy.
To see this, we first note that (\ref{ga013}) implies
\bea
\ga^0\ep_{+}=\ga^1\ep_{+}, \quad
\ga^0\chi_{-}=-\ga^1\chi_{-}
\eea
Defining $v_\pm=v_0\pm v_1$ and $\p_\pm=\p_0\pm\p_1$, and using
identities like
\bea
\ga^m\ep_{+}\p_m=\ga^1\ep_{+}\p_{+}, \quad
\ga^m\chi_{-}v_m=-\ga^1\chi_{-}v_{-}
\eea
we find after a little algebra another spinor and scalar multiplet
\bea
&& \da(\ga^1\chi_{-})=\ep_{+}v_{+}, \quad
\da v_{+}=\epbar_{+}\ga^m\p_m(\ga^1\chi_{-})
\nn\\
&& \da v_{-}=\epbar_{+}[\ga^1\la_{+}+\p_{-}\chi_{-}], \quad
\da [\ga^1\la_{+}+\p_{-}\chi_{-}]=\ga^m\ep_{+}\p_m v_{-}
\eea
We conclude that the 3D $N=1$ spinor multiplet
$\Psi_3=(\chi,M,v_\mu,\la)$ splits into the following
four 2D $N=(1,0)$ multiplets,
\bea
\label{psi32c}
&& \Psi_2=(\chi_{+}, \; M+v_3), \quad
\Phi_2=(v_3, \; \half\la_{-}+\p_3\chi_{-})
\nn\\
&& \Psi_2^\rr=(\ga^1\chi_{-}, \; v_{+}), \quad
\Phi_2^\rr=(v_{-}, \; \ga^1\la_{+}+\p_{-}\chi_{-})
\eea
Multiplying $\Psi_2^\rr$ with $\Phi_2^\rr$, we obtain a composite
2D $N=(1,0)$ spinor multiplet,
\bea
\Psi_2^\rr \times \Phi_2^\rr
=(\ga^1\chi_{-}v_{-}, \;\;
v_{+}v_{-}+\chibar_{-}\la_{+}+\chibar_{-}\ga^1\p_{-}\chi_{-})
\eea
Its highest component transforms into a total $\p_m$ derivative
under $\ep_{+}$ susy so that
\bea
\label{delCS}
2 \int_{\p\mc{M}} d^2x \Big[ v_{+}v_{-}
+\chibar_{-}\la_{+}+\chibar_{-}\ga^1\p_{-}\chi_{-} \Big]
\eea
is invariant under $\ep_{+}$ susy. Adding this boundary action
to (\ref{CS}), we obtain\footnote{
The boundary action in (\ref{CS2}) can also be written as
$-2 \int_{\p\mc{M}} d^2x (v_m v^m+\chibar_{-}\ga^m\p_m\chi_{-})$.
}
\bea
\label{CS2}
S=\int_{\mc{M}} d^3x \Big[
2 \ep^{\mu\nu\rho}v_\mu F_{\nu\rho}+\labar\la \Big] 
+2 \int_{\p\mc{M}} d^2x \Big[ v_{+}v_{-}
+\chibar_{-}\ga^1\p_{-}\chi_{-} \Big]
\eea
This bulk-plus-boundary action is invariant under $\ep_{+}$ susy.
The term linear in the auxiliary field $\la$ no longer appears 
in the boundary action so that we can eliminate it (set $\la=0$)
while preserving ``susy without BC.'' 

We observe that the elimination of the term linear in $\la$
from the boundary action has turned the hitherto pure-gauge bulk
fermionic field $\chi_{-}$ into a dynamical boundary field.
The distinctive feature of the Chern-Simons model that is responsible
for this effect is that it is gauge invariant only up to a boundary
term. In this sense it is very similar to supergravity theories
where the (super)diffeomorphism invariance of the bulk action also
holds only up to a boundary term \cite{dbpvn1}. 
Therefore, it is expected that
some of the usual pure gauge degrees of freedom (usually removed
by imposing the WZ gauge) will become important for 
bulk-plus-boundary supergravity theories.

\newpage
\section{Euler-Lagrange variation and boundary conditions} 

Our extended $F$- and $D$-term formulae give bulk-plus-boundary
actions that are ``susy without BC.'' Nevertheless, BC do arise
if one requires the Euler-Lagrange (EL) variation to vanish.
The BC one finds in this way have to be consistent with susy:
the susy variation of a given BC may generate a new BC which 
has to be added to the total set of BC, and the susy variation
of this new BC may generate yet another BC, etc.
The total set of BC forms a (finite or infinite) ``susy orbit'' 
of BC \cite{pdv,igarashi,VV}. 
In this section, using the 3D and 4D Wess-Zumino models as examples,
we show that one needs to consider only \emph{finite} susy orbits
when auxiliary fields are present. 
In the next section, we will show that such
orbits arise naturally as BC on superfields once one passes to the 
formulation in terms of co-dimension one (boundary) superfields
\cite{db1,db3}.

\subsection{3D Wess-Zumino model} 
\label{sec-3EL}

First, we simplify our equations by writing bulk-plus-boundary
actions as bulk Lagrangians with appropriate total $\p_3$ 
derivatives. The 3D Wess-Zumino bulk-plus-boundary action 
(\ref{3DWZ}) is then written as the following Lagrangian,
\bea
\mc{L}=\mc{L}_B+\p_3\mc{L}_b, \quad
\mc{L}_B=F^2-\p_\mu A\p^\mu A-\psibar\ga^\mu\p_\mu\psi, \quad
\mc{L}_b=A F+A\p_3 A
\eea
The EL variation of $\mc{L}_B$ gives
\bea
\da\mc{L}_B=(EOM)-\p_3[2\da A\p_3 A+\psibar\ga^3\da\psi]
\eea
where $(EOM)=2(F\da F+\da A\p_\mu\p^\mu A-\da\psibar\ga^\mu\p_\mu\psi)$
and we dropped an (insignificant) total $\p_m$ derivative.
For the EL variation of the total Lagrangian $\mc{L}$ we then find
\bea
\da\mc{L}=(EOM)+\p_3\Big[ A\da F+(F-\p_3 A)\da A
+A\da(\p_3 A)+\psibar_{+}\da\psi_{-}-\psibar_{-}\da\psi_{+} \Big]
\eea
Requiring this to vanish for arbitrary variations of the fields
on the boundary gives the following set of BC,
\bea
A=F-\p_3 A=\psi_{+}=\psi_{-}=0
\eea
which is obviously too strong. $(A,\p_3 A)$ and $(\psi_{-},\psi_{+})$
can be thought of as $(q,p)$ pairs of canonically conjugated 
variables with respect to the $x^3$ direction. 
Therefore, acceptable BC would be conditions on $p$ \emph{or}
$q$, but not on both of them at the same time.

For the modified action (\ref{3DWZa}), we have 
$\mc{L}_b=\psibar_{+}\psi_{-}$ so that
\bea
\label{var1}
\da\mc{L}=(EOM)+\p_3\Big[
-2(\p_3 A)\da A+2\psibar_{+}\da\psi_{-} \Big]
\eea
We see that the boundary piece of the EL variation is in the
``$p\da q$'' form, so that Neumann (N) BC ``$p=0$'' follow from
requiring $\da\mc{L}$ to vanish for arbitrary $\da q$ on the
boundary, or one can set ``$q=\text{const}$'' as Dirichlet (D) BC.
In the case at hand,\footnote{
When ``const'' stands for a multiplet (or a superfield),
it is understood that only the lowest component is a non-zero
constant, whereas higher components have to be zero by susy.
}
\bea
\label{bc1}
N: (\p_3 A, \psi_{+})=0, \quad
D: (A, \psi_{-})=\text{const}
\eea
The Dirichlet BC form a closed susy orbit, see (\ref{tr2sf}), but 
the Neuman BC, $(\p_3 A, \psi_{+})=0$, 
do not form a closed orbit.
Indeed, (\ref{tr2sf}) indicates that $(F+\p_3 A, \psi_{+})=0$
would be closed under $\ep_{+}$ susy, whereas (\ref{tr3sc})
says that omitting $F$ would lead to an infinite orbit of 
conditions involving restrictions of bulk equations of motion
to the boundary (as was observed in \cite{VV}).

However, the same action (\ref{3DWZa}) can be shown to give
rise to Neumann BC which do form a closed susy orbit.
This is achieved by a field redefinition in accordance with
the structure of the co-dimension one multiplets (\ref{cm1}).\footnote{
The necessity of such field redefinitions was discussed in \cite{db1}
for a particular 5D susy model.
}
Defining $F^\rr=F+\p_3 A$, we find that
\bea
\mc{L}_B=F^2-(\p_3 A)^2+\dots =(F^\rr)^2-2 F^\rr\p_3 A+\dots
\eea
Using $F^\rr$ as an independent bulk field gives, instead
of (\ref{var1}),
\bea
\da\mc{L}=(EOM)+\p_3\Big[
-2 F^\rr \da A+2\psibar_{+}\da\psi_{-} \Big]
\eea
and instead of (\ref{bc1}), we find the following susy orbits
of BC,
\bea
N: \Psi_2=0, \quad D: \Phi_2=\text{const}
\eea
where $\Phi_2$ and $\Psi_2$ are defined in (\ref{cm1}).
We will see later that these BC follow naturally in the
superspace formulation with co-dimension one superfields.
It then will also become obvious that if, instead of adding
(\ref{delWZ3}) to (\ref{3DWZ}), as we did to obtain (\ref{3DWZa}),
we \emph{subtract} it, the resulting bulk-plus-boundary
action would have flipped sets of BC,
\bea
N: \Phi_2=0, \quad D: \Psi_2=\text{const}
\eea

\subsection{4D Wess-Zumino model} 

The analysis of the EL variation and associated BC for the
4D Wess-Zumino model is very similar to the 3D case.
We find that the sum of (\ref{4DWZ}) and (\ref{del4DWZ}),
with $\al=\pm 1$, gives actions whose EL variations are in the
``$p\da q$'' form provided we use $F^\rr=F+\p_3 A$ and
$G^\rr=G+\p_3 B$ as independent bulk fields.
The corresponding BC are
\bea
\al=+1 \qrq N: \Phi_A=0, \quad D: \Phi_B=\text{const} \nn\\
\al=-1 \qrq N: \Phi_B=0, \quad D: \Phi_A=\text{const}
\eea
where $\Phi_A$ and $\Phi_B$ are defined in (\ref{cm2}).

\newpage
\section{Superspace approach} 

In this section we will demonstrate how the results derived so far
in the susy tensor calculus approach follow from superspace.
In particular, we will explain how co-dimension one superfields can
be obtained by projection with superspace covariant derivatives.
We will discuss only the 3D case (with 2D boundaries).\footnote{
The 4D case (with 3D boundaries) can be discussed along similar
lines but is more involved. If one chooses the original approach
to superspace due to Salam and Strathdee \cite{SS}, 
then one can keep the 4D gamma matrices
in a general representation and use them to describe co-dimension
one (3D) superfields. A more conventional approach \cite{wb,ggrs}
uses two-component spinors, which assumes a particular representation
of the 4D gamma matrices from the start.
One way to define co-dimension one superfields in this approach
was described in \cite{sakamura}. 
Their definition by projection
with covariant derivatives can also be established.
This was essentially done by Siegel in \cite{siegel},
only there the dependence of fields on extra coordinates was
suppressed.
}

\subsection{Superfields and superspace covariant derivatives} 

A superfield glues components of a susy multiplet into a single
object (a field over superspace). Using the same letter for
a multiplet and the corresponding superfield,  
the 3D $N=1$ superfields are\footnote{
Our 3D superspace conventions are close to those in
\cite{ggrs}.
}
\bea
\label{phiga}
\Phi &=& (A,\psi_\al,F)=A+\tabar\psi+\ta^2 F
\nn\\
\Ga_\al &=& (\chi_\al, M, v_\mu, \la_\al)
=\chi_\al+\ta_\al M+(\ga^\mu\ta)_\al v_\mu+\ta^2
\Big[ \la_\al-(\ga^\mu\p_\mu\chi)_\al \Big]
\eea
where $\ta_\al$ is an anticommuting parameter 
(a two-component 3D Majorana spinor) and
\bea
\ta^2=\half\tabar\ta=\half \ta^T C\ta
=\half\ta_\al C^{\al\beta}\ta_\beta
=\half\ta^\al\ta_\al
\eea
Here we introduced spinor indices $\al$ that so far
have been hidden in our notation. Keeping these indices
explicit is often convenient in superspace calculations.
In our conventions,
\bea
& \tabar\psi=\ta^\al\psi_\al, \quad
(\ga^\mu\ta)_\al=(\ga^\mu)_\al{}^\beta\ta_\beta, \quad
\ta^\al=\ta_\beta C^{\beta\al}, \quad
\ta_\al=\ta^\beta C_{\beta\al}, \quad
\ta_\al\ta_\beta=-C_{\al\beta}\ta^2 &
\nn\\
& C_{\al\beta}C^{\beta\ga}=\da_\al{}^\ga, \quad
C_{\al\beta}=-C_{\beta\al}, \quad
\ga^\mu_{\al\beta} \equiv (\ga^\mu)_\al{}^\ga C_{\ga\beta}
=\ga^\mu_{\beta\al} &
\eea
Susy transformations of superfields are generated by 
differential operators $Q_\al$,
\bea
\label{defQ}
\da \Phi=\epbar Q\Phi, \quad
\da \Ga_\al=\epbar Q\Ga_\al; \quad
Q_\al=\p_\al-(\ga^\mu\ta)_\al \p_\mu, \quad
\p_\al=\frac{\p}{\p\ta^\al}
\eea
On the component level, this gives the transformations
(\ref{tr3sc}) and (\ref{tr3sp}).
In our conventions, $\p_\al\ta^\beta=\da_\al{}^\beta$
and $\p_\al\ta_\beta=C_{\al\beta}$, so that introducing
\bea
D_\al=\p_\al+(\ga^\mu\ta)_\al \p_\mu
\eea
we obtain the following algebra
\bea
\{ Q_\al, Q_\beta \} =2\ga^\mu_{\al\beta}\p_\mu, \quad
\{ Q_\al, D_\beta \} =0, \quad
\{ D_\al, D_\beta \} =-2\ga^\mu_{\al\beta}\p_\mu
\eea
The second property, $\{ Q,D\}=0$, implies that $D_\al$ are
superspace covariant derivatives,
\bea
\da (D_{\al_1} \dots D_{\al_n} \Phi)=\epbar Q
(D_{\al_1} \dots D_{\al_n} \Phi)
\eea
For example, $D_\al\Phi$ is a spinor multiplet like $\Ga_\al$.
This is used to define superfield gauge transformations as
\bea
\label{gtr}
\da_g \Ga_\al=(\da\chi_\al, \; \da M, \; \da v_\mu, \; \da\la_\al)
=D_\al\Phi=(\psi_\al, \; F, \; \p_\mu A, \; 0)
\eea
so that $v_\mu$ transforms like a gauge field 
and $\la_\al$ is gauge-invariant.
When such a superfield transformation is a symmetry of the action,
one can impose a Wess-Zumino gauge: $\chi_\al=M=0$.

As the indices $\al$ are two-dimensional, 
$[D_\al,D_\beta]$ is proportional to $C_{\al\beta}$
and we find 
\bea
\label{DD3D}
D_\al D_\beta=-\ga^\mu_{\al\beta}\p_\mu-C_{\al\beta}D^2, \quad
D^2=\half D^\al D_\al
\eea
As the complete antisymmetrization of three two-dimensional indices 
gives zero, we find the following identity
\bea
D_\al D_\beta D_\ga=\half D_\al\{D_\beta,D_\ga\}
-\half D_\beta\{D_\al,D_\ga\}
+\half D_\ga\{D_\al,D_\beta\}
\eea
It then follows that an arbitrary product of $D_\al$ can be
written as a linear combination of $1$, $D_\al$ and $D^2$ with
$\p_\mu$-dependent coefficients. For example,
\bea
\label{DnD}
D^\al D_\beta D_\al=0, \quad
D^2 D_\al=-D_\al D^2=(\ga^\mu D)_\al \p_\mu, \quad
D^2 D^2=\p_\mu \p^\mu
\eea
In turn, this implies that all the independent components of a 
3D $N=1$ superfield $S$ can be defined in terms of lowest components
of $S$, $D_\al S$ and $D^2 S$. For example,
\bea
\label{cobypr}
\Phi &=& (A, \psi_\al, F)=(\Phi, \;\; D_\al\Phi, \;\; -D^2\Phi)_| 
\nn\\
\Ga_\al &=& (\chi_\al, M, v_\mu, \la_\al)=\Big(
\Ga_\al, \;\; -\half \ov D\Ga, \;\;
-\half \ov D \ga^\mu\Ga, \;\;
-D^2\Ga_\al+(\ga^\mu\p_\mu\Ga)_\al \Big)_|
\eea
where the bar ``$|$'' indicates setting $\ta=0$.
The fact that $\la_\al$ is a gauge-invariant component field
corresponds to the fact that
\bea
\label{defw}
w_\al=-D^\beta D_\al\Ga_\beta
=-D^2\Ga_\al+(\ga^\mu\p_\mu\Ga)_\al
\eea
is a gauge-invariant superfield. We find (compare with $\Ga_\al$
in (\ref{phiga})),
\bea
w_\al=(\la_\al, \; 0, \; 
-\ep_{\mu\nu\rho}F^{\nu\rho}, \; 2(\ga^\mu\p_\mu\la)_\al)\
=\la_\al+(\ga^{\mu\nu}\ta)_\al F_{\mu\nu}+\ta^2(\ga^\mu\p_\mu\la)_\al
\eea
Note that $-D^2\Phi$ and $w_\al$
correspond to the kinetic multiplets (\ref{Tphi3}) and (\ref{Tpsi3}),
respectively.

\subsection{Co-dimension one superfields} 

We now proceed to decompose the 3D $N=1$ superfields
$\Phi$ and $\Ga_\al$ into 2D $N=(1,0)$ superfields transforming 
in the standard way under $\ep_{+}$ susy. First, we write
\bea
\epbar Q=\epbar_{+}Q_{-}+\epbar_{-}Q_{+}, \quad
\ep_\pm=P_\pm\ep, \quad
Q_\pm \equiv P_\pm Q
\eea
where $P_\pm=\half(1\pm\ga^3)$. 
From (\ref{defQ}), using $\mu=(m,3)$, 
we obtain
\bea
Q_{-}=Q_{-}^\rr+\ta_{-}\p_3, \quad
Q_{-\al}^\rr \equiv \p_{-\al}-(\ga^m\ta_{+})_\al \p_m, \quad
\p_{-\al} \equiv \frac{\p}{\p\ta_{+}^\al}
\nn\\
Q_{+}=Q_{+}^\rr-\ta_{+}\p_3, \quad
Q_{+\al}^\rr \equiv \p_{+\al}-(\ga^m\ta_{-})_\al \p_m, \quad
\p_{+\al} \equiv \frac{\p}{\p\ta_{-}^\al}
\eea
By definition, $Q_{-}$ is the generator of
$\ep_{+}$ susy transformations on 3D $N=1$ superfields,
\bea
\da_{+}\Phi=(\epbar_{+}Q_{-})\Phi, \quad
\da_{+}\Ga_\al=(\epbar_{+}Q_{-})\Ga_\al
\eea
On the other hand, $Q_{-}^\rr$ has the standard form for the generator
of $\ep_{+}$ susy transformations on 2D $N=(1,0)$ superfields.
The two operators are related as follows (as was also observed
and used in \cite{sakai}),
\bea
Q_{-}^\rr=Q_{-}-\ta_{-}\p_3=
e^{+\tabar_{+}\ta_{-}\p_3}Q_{-}
e^{-\tabar_{+}\ta_{-}\p_3}
\eea
Therefore, writing
\bea
\label{phi32}
\boxed{ \rule[-5pt]{0pt}{16pt} \quad
\Phi=e^{-\tabar_{+}\ta_{-}\p_3}\Big[
\wh A+\tabar_{-}\wh\psi_{+} \Big]
\quad}
\eea
we find
\bea
\da_{+}\Phi=(\epbar_{+}Q_{-})\Phi
=e^{-\tabar_{+}\ta_{-}\p_3}\Bigg\{ \epbar_{+}Q_{-}^\rr
\Big[ \wh A+\tabar_{-}\wh\psi_{+} \Big] \Bigg\}
=e^{-\tabar_{+}\ta_{-}\p_3}\Big[
(\da_{+}\wh A)+\tabar_{-}(\da_{+}\wh\psi_{+}) \Big]
\eea
so that the $\ta_{+}$-dependent objects $\wh A$ and $\wh\psi_{+}$
defined by (\ref{phi32}) are, indeed, 2D $N=(1,0)$ superfields.
The $\ta_{+}$ expansions of these superfields follow from (\ref{phi32}),
\bea
\wh A+\tabar_{-}\wh\psi_{+}=e^{+\tabar_{+}\ta_{-}\p_3}\Phi
=(1+\tabar_{+}\ta_{-}\p_3)(A+\tabar_{+}\psi_{-}+\tabar_{-}\psi_{+}
+\tabar_{+}\ta_{-}F)
\eea
which gives\footnote{
We denote the co-dimension one superfields by the same letter
as the corresponding lowest component, but with a hat on it.
These lowest components can be obtained by setting $\ta_{+}=0$
in the 3D superfield, e.g. $\Phi(\ta_{+}=0)=A+\tabar_{-}\psi_{+}$.
}
\bea
\boxed{ \rule[-5pt]{0pt}{16pt} \quad
\wh A=A+\tabar_{+}\psi_{-}, \quad
\wh\psi_{+}=\psi_{+}+\ta_{+}(F+\p_3 A)
\quad}
\eea
The superfield transformations
$\da_{+}\wh A=(\epbar_{+}Q_{-}^\rr)\wh A$ and
$\da_{+}\wh\psi_{+}=(\epbar_{+}Q_{-}^\rr)\wh\psi_{+}$
give rise to the component susy transformations (\ref{tr2sf}).

The co-dimension one superfields can also be defined by projection
with superspace covariant derivatives. To this extent, we 
decompose $D_\al$ into $D_{\pm\al}=(P_{\pm}D)_\al$,
\bea
D_{-}=D_{-}^\rr-\ta_{-}\p_3, \quad
D_{-\al}^\rr \equiv \p_{-\al}+(\ga^m\ta_{+})_\al \p_m
\nn\\
D_{+}=D_{+}^\rr+\ta_{+}\p_3, \quad
D_{+\al}^\rr \equiv \p_{+\al}+(\ga^m\ta_{-})_\al \p_m
\eea
and observe that
\bea
D_{+}^\rr=D_{+}-\ta_{+}\p_3=
e^{+\tabar_{+}\ta_{-}\p_3}D_{+}
e^{-\tabar_{+}\ta_{-}\p_3}
\eea
Acting with $D_{+\al}$ on $\Phi$ and setting $\ta_{-}=0$ then gives
\bea
D_{+}\Phi_{|\ta_{-}=0}=D_{+}^\rr\Big[
\wh A+\tabar_{-}\wh\psi_{+} \Big]{}_{|\ta_{-}=0}=\wh\psi_{+}
\eea
where we used that
\bea
\p_{+\al}\ta_{-}^\beta=(P_{+})_\al{}^\ga\p_\ga\ta^\da (P_{+})_\da{}^\beta
=(P_{+})_\al{}^\ga(P_{+})_\ga{}^\beta
=(P_{+})_\al{}^\beta
\eea
As a result, the co-dimension one decomposition of $\Phi$ by
projection is given by
\bea
\label{sfbypr}
\boxed{ \rule[-5pt]{0pt}{16pt} \quad
\wh A=\Phi_{|\ta_{-}=0}, \quad
\wh\psi_{+}=D_{+}\Phi_{|\ta_{-}=0}
\quad}
\eea

The decomposition of the 3D $N=1$ spinor multiplet $\Ga_\al$
is quite similar. We find,
\bea
\Ga_{+} &=& e^{-\tabar_{+}\ta_{-}\p_3}\Big[
\wh\chi_{+}-\ga^1\ta_{-}\wh v_{-} \Big]
\nn\\
\Ga_{-} &=& e^{-\tabar_{+}\ta_{-}\p_3}\Big[
\wh\chi_{-}+\ta_{-}(\wh M-\wh v_3) \Big]
\eea
where
\bea
\wh\chi_{+} &=& \chi_{+}+\ta_{+}(M+v_3), \quad
\wh v_{-} = v_{-}+\tabar_{+}\ga^1[\la_{+}+\ga^1\p_{-}\chi_{-}] 
\nn\\
\wh\chi_{-} &=& \chi_{-}+\ga^1\ta_{+}v_{+}, \quad
(\wh M-\wh v_3) = (M-v_3)
-\tabar_{+}[\la_{-}-\ga^1\p_{+}\chi_{+}+2\p_3\chi_{-}]
\eea
Observing that $-\ov D_{-}^\rr\wh\chi_{+}
=M+v_3+\tabar_{+}\ga^1\p_{+}\chi_{+}$, we further find
\bea
\wh M=M+\tabar_{+}\Big[
-\half\la_{-}+\ga^1\p_{+}\chi_{+}-\p_3\chi_{-} \Big], \quad
\wh v_3=v_3+\tabar_{+}\Big[
\half\la_{-}+\p_3\chi_{-} \Big]
\eea
The multiplets $\wh\chi_{+}$, $\wh\chi_{-}$, $\wh v_{-}$ and $\wh v_3$
match those in (\ref{psi32c}). These multiplets can also be defined
by projection. For the following, we only note that
\bea
\label{prGa}
\Ga_{-}{}_{|\ta_{-}=0}=\wh\chi_{-}, \quad
(D_{+\al}\Ga_{+\beta})_{|\ta_{-}=0}=(P_{+}\ga^1)_{\al\beta}\wh v_{-}
\eea

\newpage
\subsection{Co-dimension one decomposition of 3D Lagrangians} 

In 3D, an $N=1$ susy Lagrangian is usually defined as 
the $F$-term of a scalar superfield,
\bea
\mc{L}=F=[\Phi]_F=\int d^2\ta\Phi=-D^2\Phi_|
\eea
Such a Lagrangian transforms into a total $\p_\mu=(\p_m,\p_3)$ 
derivative and is not susy in the presence of a boundary.
Using the following identity (that will be proven shortly),
\bea
\label{D2pm3}
D^2=\ov D_{-} D_{+}+\p_3
\eea
we find that the following modified Lagrangian,
\bea
\label{modL}
\mc{L}^\rr=F+\p_3 A=[\Phi]_F+\p_3(\Phi_|)=-\ov D_{-} D_{+}\Phi{}_|
=-\ov D_{-}^\rr\wh\psi_{+}{}_{|\ta_{+}=0}=[\wh\psi_{+}]_f
\eea
is written as the $f$-term of a 2D $N=(1,0)$ spinor superfield
$\wh\psi_{+}=\psi_{+}+\ta_{+}f$. Therefore, under $\ep_{+}$ susy,
it transforms into a total $\p_m$ derivative and is susy in the
presence of a boundary at $x^3=\text{const}$.
This way we recover our ``$F+A$'' formula (\ref{F3}) and also
obtain a way to rewrite the resulting modified Lagrangian in terms
of co-dimension one superfields.

To prove (\ref{D2pm3}), we first project (\ref{DD3D}) with $P_{\pm}$
to find that
\bea
D_{-\al}D_{+\beta} &=& (P_{-})_\al{}^\ga (P_{+})_\beta{}^\da D_\ga D_\da
\nn\\
&=& -(P_{-}\ga^\mu P_{-})_{\al\beta}\p_\mu-(P_{-}P_{-})_{\al\beta} D^2
=(P_{-})_{\al\beta}(\p_3-D^2)
\eea
where we used $(P_{-})_{\al\beta}=-(P_{+})_{\beta\al}$ as follows
from $(P_{\pm})_{\al\beta}=\half(C_{\al\beta}\pm\ga^3_{\al\beta})$.
Contraction with $C^{\al\beta}$ gives
\bea
C^{\al\beta}D_{-\al}D_{+\beta}=\ov D_{-} D_{+}
=-(P_{-})_\al{}^\al(\p_3-D^2)=-(\p_3-D^2)
\eea
which proves (\ref{D2pm3}). Altogether, (\ref{DD3D}) decomposes as
\bea
D_{-\al}D_{-\beta}=-(\ga^m P_{+})_{\al\beta}\p_m, \quad
D_{-\al}D_{+\beta}=(P_{-})_{\al\beta}(\p_3-D^2)
\nn\\
D_{+\al}D_{+\beta}=-(\ga^m P_{-})_{\al\beta}\p_m, \quad
D_{+\beta}D_{-\al}=(P_{-})_{\al\beta}(\p_3+D^2)
\eea
from which we find that
\bea
\{ D_{\pm\al}, D_{\pm\beta} \}=-2(P_{\pm}\ga^m )_{\al\beta}\p_m, \quad
\{ D_{-\al}, D_{+\beta} \}=2(P_{-})_{\al\beta}\p_3
\eea
Now we are ready to apply the formalism to specific examples.

\newpage
\subsection{3D Wess-Zumino model} 

We start with the 3D Lagrangian,
\bea
\label{form1}
\mc{L}=D^2(\Phi D^2\Phi)_|
=F^2-\psibar\ga^\mu\p_\mu\psi+A\p_\mu\p^\mu A
\eea
The modified Lagrangian (\ref{modL}) is given by
\bea
\mc{L}^\rr=\mc{L}+\p_3(A F)=D_{-}^\al D_{+\al}(\Phi D^2\Phi)_|
\eea
and corresponds to the bulk-plus-boundary action (\ref{3DWZ}).
To write this Lagrangian in terms of co-dimension one superfields,
we have to move the $D_{+\al}$ past all $D_{-}$ and then set
$\ta_{-}=0$. Using
\bea
\label{DDp3}
D^2=\ov D_{-} D_{+}+\p_3, \quad
D_{+\al}D^2=-(\ga^m D_{-})_\al \p_m -D_{+\al}\p_3
\eea
(the second identity follows from (\ref{DnD}) by projection),
we find
\bea
\mc{L}^\rr=D_{-}^\al\Big[
(D_{+\al}\Phi)\p_3\Phi 
-\Phi\p_3(D_{+\al}\Phi)
+(D_{+\al}\Phi)(\ov D_{-} D_{+}\Phi)
-\Phi(\ga^m D_{-})_\al \p_m\Phi \Big]_|
\eea
Setting $\ta_{-}=0$ gives
\bea
\mc{L}^\rr=D_{-}^{\rr\al}\Big[
\wh\psi_{+\al}\p_3\wh A
-\wh A\p_3\wh\psi_{+\al}
+\wh\psi_{+\al}(\ov D_{-}^\rr \wh\psi_{+})
-\wh A(\ga^m D_{-}^\rr)_\al \p_m\wh A \Big]_{|\ta_{+}=0}
\eea
This Lagrangian is written in terms of 2D $N=(1,0)$ superfields
and is manifestly $\ep_{+}$ susy (it varies into a total $\p_m$
derivative) in the presence of a boundary at $x^3=\text{const}$.
The EL variation, on the other hand, gives
\bea
\da\mc{L}^\rr=(EOM)+\p_3 \Big\{ D_{-}^{\rr\al}\Big[
\wh\psi_{+\al}\da\wh A
-\wh A \da\wh\psi_{+\al} \Big]_{|\ta_{+}=0} \Big\}
\eea
We observe that $\wh A$ and $\wh\psi_{+}$ are conjugated superfields,
with respect to the ``time derivative''~$\p_3$,
but the boundary variation is not in the ``$p\da q$'' form.
It is however easy to see which separately susy boundary Lagrangians
can be added to bring the boundary piece of the EL variation
to the ``$p\da q$'' form. Defining
\bea
\mc{L}_{\pm}^\rr=\mc{L}^\rr \pm \p_3 \Delta, \quad
\Delta=D_{-}^{\rr\al}[\wh\psi_{+\al}\wh A]_{|\ta_{+}=0}
=-A(F+\p_3 A)+\psibar_{-}\psi_{+}
\eea
we find that the boundary piece of the EL variation and the
corresponding Neumann (N) and Dirichlet (D) boundary conditions are
\bea
\label{sfbc1}
\ba{cccccrcr}
\da\mc{L}_{+}^\rr &\qrq& 2\wh\psi_{+}\da\wh A &
\qrq& N: & \wh\psi_{+}=0, & \quad D: & \wh A=\text{const}
\\[5pt]
\da\mc{L}_{-}^\rr &\qrq& -2\wh A \da\wh\psi_{+} &
\qrq& N: & \wh A=0, & \quad D: & \wh\psi_{+}=\text{const}
\ea
\eea
The boundary Lagrangian $\Delta$ corresponds to the one in (\ref{delWZ3}).

Instead of (\ref{form1}), one could start with an alternative
3D Lagrangian,
\bea
\mc{L}_2=-D^2 \Big( \half D^\al\Phi D_\al\Phi \Big)_|
=F^2-\psibar\ga^\mu\p_\mu\psi-\p_\mu A\p^\mu A
\eea
that differs from (\ref{form1}) by a total $\p_\mu$ derivative.
The modified Lagrangian (\ref{modL}) is now
\bea
\mc{L}_2^\rr=\mc{L}_2+\p_3\Big( \half\psibar\psi \Big)
=-\ov D_{-}^\al D_{+\al}\Big(
\ov D_{-}\Phi D_{+}\Phi \Big)_|
\eea
which in terms of co-dimension one superfields becomes
\bea
\mc{L}_2^\rr=D_{-}^{\rr\al}\Big[
2\wh\psi_{+\al}\p_3\wh A
+(D_{-}^{\rr\beta}\wh\psi_{+\al})\wh\psi_{+\beta}
+\p_m\wh A(\ga^m D_{-}^\rr)_\al\wh A \Big]_{|\ta_{+}=0}
\eea
This way we get directly a Lagrangian whose boundary piece
of the EL variation is in the ``$p\da q$'' form. 
One can check that $\mc{L}_2^\rr$ differs from $\mc{L}_{+}^\rr$
by an (insignificant) total $\p_m$ derivative.

Adding a superpotential would not change the form of the 
superfield boundary conditions. To see this, let us consider
\bea
\mc{L}_3=-D^2\Big[ W(\Phi) \Big]_|
=-\half W^{\rr\rr}(A)\psibar\psi+W^\rr(A)F
\eea
The modified Lagrangian (\ref{modL}) is\footnote{
The fact that the bulk superpotential $W(A)$ is a natural
boundary Lagrangian was observed in \cite{warner}.
In~\cite{LRN} this was also derived using superspace methods,
but the general philosophy of that work was to use BC for
susy. Here we emphasize that the co-dimension one superspace
methods give rise to bulk-plus-boundary actions that are
``susy without BC.''
}
\bea
\mc{L}_3^\rr=\mc{L}_3+\p_3\Big[ W(A) \Big]
=-\ov D_{-} D_{+} \Big[ W(\Phi) \Big]_|
=D_{-}^{\rr\al}\Big[ -W(\wh A)\wh\psi_{+\al} \Big]_{|\ta_{+}=0}
\eea
Obviously, adding this to $\mc{L}_{\pm}^\rr$ would not change
the BC (\ref{sfbc1}). However, on the component level, one
could look for the form of BC with the auxiliary field $F$ eliminated.
Then the superpotential $W$ would explicitly appear in the BC
as in that case $2 F=-W^\rr(A)$.

\subsection{3D Chern-Simons model} 

The superfield 3D Lagrangian for the Chern-Simons model is
\bea
\mc{L}=-D^2(\ov w \Ga)_|
=2\ep^{\mu\nu\rho}v_\mu F_{\nu\rho}
+\labar\la+\p_\mu(\chibar\ga^\mu\la)
\eea
The modified Lagrangian (\ref{modL}) is
\bea
\mc{L}^\rr=\mc{L}+\p_3(\labar\chi)
=-\ov D_{-}^\al \Big[ D_{+\al} \ov w\Ga \Big]_|
\eea
Using (\ref{defw}) and (\ref{DDp3}), we find that
\bea
D_{+\al}(\ov w\Ga) &=& 
D_{+\al}(-\ov\Ga D^2\Ga+\ov\Ga\ga^\mu\p_\mu\Ga) \nn\\
&=& -(D_{+\al}\ov\Ga)D^2\Ga
+\ov\Ga (D_{+\al}D^2\Ga)
+(D_{+\al}\ov\Ga)\ga^\mu\p_\mu\Ga
-\ov\Ga\ga^\mu\p_\mu(D_{+\al}\Ga) \nn\\
&=& -(D_{+\al}\ov\Ga)\p_3\Ga
-\ov\Ga \p_3(D_{+\al}\Ga)
+(D_{+\al}\ov\Ga)\ga^3\p_3\Ga
-\ov\Ga\ga^3\p_3(D_{+\al}\Ga)+(\text{no }\p_3) \nn\\
&=& -2(D_{+\al}\ov\Ga_{+})\p_3\Ga_{-}
-2\ov\Ga_{-} \p_3(D_{+\al}\Ga_{+})+(\text{no }\p_3)
\eea
where we dropped terms not involving $\p_3$. As a result,
\bea
\mc{L}^\rr=2\ov D_{-}^\al\Big[
(D_{+\al}\ov\Ga_{+})\p_3\Ga_{-}+\ov\Ga_{-}\p_3(D_{+\al}\Ga_{+})
+(\text{no } \p_3) \Big]_|
\eea
Setting $\ta_{-}=0$ and using (\ref{prGa}), we arrive at
\bea
\mc{L}^\rr=2 D_{-}^{\rr\al}\Big[
\wh v_{-}\p_3(\ga^1\wh\chi_{-})_\al 
-(\ga^1\wh\chi_{-})_\al \p_3\wh v_{-} 
+(\text{no } \p_3) \Big]_{|\ta_{+}=0}
\eea
This shows that $\wh v_{-}$ and $\wh\chi_{-}$ are the conjugated
co-dimension one superfields for the Chern-Simons model.
Again, we can define two Lagrangians for which the boundary piece
of the EL variation is in the ``$p\da q$'' form,
\bea
\mc{L}_{\pm}^\rr=\mc{L}^\rr \pm 2\p_3 \Delta, \quad
\Delta=D_{-}^{\rr\al}\Big[
\wh v_{-}(\ga^1\wh\chi_{-})_\al \Big]_{|\ta_{+}=0}
=v_{+}v_{-}+\chibar_{-}\la_{+}+\chibar_{-}\ga^1\p_{-}\chi_{-}
\eea
The boundary piece of the EL variation and the
superfield Neumann and Dirichlet BC for these Lagrangians are as follows,
\bea
\label{sfbc2}
\ba{cccccrcr}
\da\mc{L}_{+}^\rr &\qrq& 4\wh v_{-}\da(\ga^1\wh\chi_{-}) &
\qrq& N: & \wh v_{-}=0, & \quad D: & \wh\chi_{-}=\text{const}
\\[5pt]
\da\mc{L}_{-}^\rr &\qrq& -4(\ga^1\wh\chi_{-}) \da\wh v_{-}  &
\qrq& N: & \wh\chi_{-}=0, & \quad D: & \wh v_{-}=\text{const}
\ea
\eea
The boundary Lagrangian $\Delta$ corresponds to the one in (\ref{delCS}).

Note that deriving these BC in the component formulation is tricky
as one has to choose appropriate independent bulk fields
(namely, $\la_{-}^\rr=\la_{-}-\ga^1\p_{+}\chi_{+}+2\p_3\chi_{-}$)
as dictated by the way fields appear in the co-dimension one
superfields.

\subsection{3D Maxwell model} 

The superfield 3D Lagrangian for the Maxwell model is
\bea
\mc{L}=-D^2(\ov w w)_|
=4 F_{\mu\nu} F^{\mu\nu} +2\labar\ga^\mu\p_\mu\la
\eea
The modified Lagrangian (\ref{modL}) is
\bea
\mc{L}^\rr=\mc{L}+\p_3(\labar\la)
=-D_{-}^\al D_{+\al}(2\ov w_{+} w_{-})_|
\eea
where the projections $w_{\pm}$, as follows from (\ref{defw}), are
\bea
w_{+} &=& \ga^m\p_m\Ga_{-}-\ov D_{-} D_{+}\Ga_{+} \nn\\
w_{-} &=& \ga^m\p_m\Ga_{+}-\ov D_{-} D_{+}\Ga_{-}-2\p_3\Ga_{-}
\eea
To find conjugated co-dimension one superfields in this model,
we perform the co-dimension one decomposition of
the EL variation $\da\mc{L}^\rr$ and look for terms with $\p_3$
acting on variations of superfields. 
Using $D_{+\al}\ov D_{-} D_{+}=-2\p_3 D_{+\al}+(\text{no }\p_3)$,
we find that
\bea
D_{+\al}(\ov w_{+}\da w_{-}+\ov w_{-}\da w_{+})
=-2(D_{+\al}\ov w_{+})\p_3\da\Ga_{-}
-2\ov w_{-}\p_3(D_{+\al}\Ga_{+})
+(\text{no }\p_3\da\Ga)
\eea
Therefore, the EL variation of $\mc{L}^\rr$ reads
\bea
\label{daM}
\da\mc{L}^\rr=(EOM)+4\p_3\Big\{ D_{-}^\al \Big[
\ov w_{-}\da(D_{+\al}\Ga_{+})+(D_{+\al}\ov w_{+})\da\Ga_{-} \Big]_|
\Big\}
\eea
This shows that, unlike the Wess-Zumino and Chern-Simons models,
here we have \emph{two pairs} of conjugated co-dimension one superfields
and the EL variation is already in the ``$p\da q$'' form.

To write this more explicitly, we need an analog of (\ref{prGa})
for $w_\al$. First, we find that
\bea
w_{+} &=& \la_{+}+\ta_{+} F_{+-}+2\ga^1\ta_{-} F_{-3}
+\tabar_{+}\ta_{-}(-\ga^1\p_{-}\la_{-}+\p_3\la_{+}) \nn\\
w_{-} &=& \la_{-}-\ta_{-} F_{+-}+2\ga^1\ta_{+} F_{+3}
+\tabar_{+}\ta_{-}(\ga^1\p_{+}\la_{+}-\p_3\la_{-})
\eea
where $F_{+-}=\p_{+}v_{-}-\p_{-}v_{+}$,
$F_{+3}=\p_{+}v_3-\p_3 v_{+}$,
$F_{-3}=\p_{-}v_3-\p_3 v_{-}$
(or, equivalently,
$F_{+-}=-2 F_{01}$,
$F_{+3}=F_{03}+F_{13}$,
$F_{-3}=F_{03}-F_{13}$) with 
$v_\pm=v_0\pm v_1$ and
$\p_\pm=\p_0\pm\p_1$. 
This leads to the following decomposition,
\bea
w_{+} &=& e^{-\tabar_{+}\ta_{-}\p_3}\Big[
\wh\la_{+}+2\ga^1\ta_{-}\wh F_{-3} \Big]
\nn\\
w_{-} &=& e^{-\tabar_{+}\ta_{-}\p_3}\Big[
\wh\la_{-}-\ta_{-}\wh F_{+-} \Big]
\eea
where
\bea
&& \wh\la_{+}=\la_{+}+\ta_{+} F_{+-}, \quad
\wh F_{-3}=F_{-3}+\half\tabar_{+}(\p_{-}\la_{-}-2\ga^1\p_3\la_{+}) \nn\\
&& \wh\la_{-}=\la_{-}+2\ga^1\ta_{+} F_{+3}, \quad
\wh F_{+-}=F_{+-} +\tabar_{+}\ga^1\p_{+}\la_{+}
\eea
These superfields can also be defined by projection.
We only need two of the projections,
\bea
w_{-} {}_{|\ta_{-}=0}=\wh\la_{-}, \quad
(D_{+\al}w_{+\beta})_{|\ta_{-}=0} 
=-2(P_{+}\ga^1)_{\al\beta} \wh F_{-3}
\eea
Together with (\ref{prGa}), this allows us to rewrite (\ref{daM}) as
\bea
\da\mc{L}^\rr=(EOM)+4\p_3\Big\{ -D_{-}^{\rr\al} \Big[
(\ga^1\wh\la_{-})_\al \da\wh v_{-}
+2\wh F_{-3}(\ga^1\da\wh\chi_{-}) \Big]_{|\ta_{+}=0} \Big\}
\eea
This clearly shows $(\wh\la_{-}, \wh v_{-})$ and
$(\wh F_{-3}, \wh\chi_{-})$ as the two pairs of conjugated
co-dimension one superfields. (In components, we have
\bea
\da\mc{L}^\rr &=& (EOM)+4\p_3\Big\{
2 F_{+3}\da v_{-}+2 F_{-3}\da v_{+} 
\nn\\ &&\hspace{80pt}
+\labar_{-}\da(\la_{+}+\ga^1\p_{-}\chi_{-})
+\da\chibar_{-}(\ga^1\p_{-}\la_{-}-2\p_3\la_{+}) \Big\}
\eea
Proving this on the component level is rather tricky, as one
has to define $\la_{+}^\rr=\la_{+}+\ga^1\p_{-}\chi_{-}$
and $\la_{-}^\rr=\la_{-}-\ga^1\p_{+}\chi_{+}+2\p_3\chi_{-}$
and consider them as independent bulk fields.)

In the Maxwell model, we can define \emph{four} Lagrangians with
different sets of BC. Namely,
\bea
\mc{L}_1^\rr=\mc{L}^\rr, \quad
\mc{L}_2^\rr=\mc{L}^\rr+4\p_3\Delta_1, \quad
\mc{L}_3^\rr=\mc{L}^\rr+4\p_3\Delta_2, \quad
\mc{L}_4^\rr=\mc{L}^\rr+4\p_3(\Delta_1+\Delta_2)
\eea
with
\bea
\Delta_1=D_{-}^{\rr\al}\Big[ 
(\ga^1\wh\la_{-})_\al \wh v_{-} \Big]_{|\ta_{+}=0}, \quad
\Delta_2=D_{-}^{\rr\al}\Big[ 
2\wh F_{-3}(\ga^1\wh\chi_{-})_\al \Big]_{|\ta_{+}=0}
\eea
The Neumann BC in the four cases are, respectively,
\bea
(\wh\la_{-}, \wh F_{-3})=0, \quad
(\wh v_{-}, \wh F_{-3})=0, \quad
(\wh\la_{-}, \wh\chi_{-})=0, \quad
(\wh v_{-}, \wh\chi_{-})=0
\eea
Each of these four sets of BC is closed under $\ep_{+}$ susy.
The first set is also gauge-invariant. 

\section{Conclusions}

In this article we have made a systematic study of boundary
conditions (BC) in rigidly supersymmetric (susy) models.
We first analyzed the models in $x$-space, and were able to
construct susy bulk-plus-boundary actions which were susy by
themselves,
without the need for BC. We called such actions ``susy without BC.''
To achieve this, we had to add boundary actions which completed
the bulk actions, but which themselves were not susy.
In some cases we ended up with models which contained boundary
terms which were linear in auxiliary fields. Since elimination
of auxiliary fields in such models gave too strong BC, we added
separately susy actions on the boundary which canceled the terms
linear in auxiliary fields.

In the tensor calculus approach, the key to the construction
of susy bulk-plus-boundary actions was our extended $F$-term
formula (or ``$F+A$'' formula): (\ref{F3}) in 3D and (\ref{F4})
in 4D. In 4D, we found also an extended $D$-term
formula (\ref{D4}). For constructing separately susy boundary
actions, we needed in addition to decompose bulk susy multiplets
into a set of co-dimension one multiplets out of which, using
standard tensor calculus methods, we could construct susy boundary
actions. 

To construct the susy bulk-plus-boundary actions in superspace 
(which we discussed explicitly only for the 3D case),
we used the decomposition in (\ref{D2pm3}),
\bea
D^2=\ov D_{-} D_{+}+\p_3
\eea
where $D=D_\al$ are the usual superspace covariant derivatives
(with the spin index $\al$),
and $D_{\pm}=P_{\pm}D$ with $P_{\pm}=\half(1\pm\ga^3)$.
The modified Lagrangian 
$\mc{L}^\rr=(-D^2+\p_3)\Phi_|$ for a composite superfield
$\Phi$ consisted of the usual bulk term $F$ from $-D^2\Phi_|$, and the 
boundary term $A$ from $\p_3\Phi_|$ which is to be added on the 
boundary. So, starting from the Lagrangian 
$\mc{L}^\rr=-\ov D_{-} D_{+}\Phi_|=-D_{-}^\al D_{+\al}\Phi_|$, 
the nonsupersymmetric
boundary term ``$A$'' which completes the bulk action ``$F$''
is included from the start.

The operators $D_{+}=D_{+\al}$ were used to decompose a bulk superfield
which depends on $\ta_{+}$ and $\ta_{-}$
into a set of co-dimension one superfields 
which depend only on $\ta_{+}$.
While the components of a superfield are defined by acting on it
with $D_\al$ and setting $\ta_{+}=\ta_{-}=0$, see (\ref{cobypr}),
we defined the co-dimension one superfields by acting on the
parent superfield
with $D_{+\al}$ and setting $\ta_{-\al}=0$, see (\ref{sfbypr}).
This approach led naturally to the foliation of bulk superfields into
co-dimension one (boundary) superfields which is
similar to the decomposition
of $N=2$ superfields into $N=1$ superfields.
Using these co-dimension one superfields we could construct
separately susy boundary actions using the usual superspace
methods. The susy covariant
derivatives $D_{-}^\rr$ of the lower-dimensional superspace
(which depend on $\ta_{+}$ and~$\p_m$, but not on $\ta_{-}$
and $\p_3$) were obtained by setting $\ta_{-}=0$ in $D_{-}$.

We conclude that the component approach and the superspace
approach remain equivalent in the presence of boundaries.

An issue we want now to confront concerns the BC for Euler-Largange
(EL) variations. In various cases we were able to add separately susy 
boundary actions such that the EL variation of the action was
of the form ``$p\da q$'' on the boundary. We thus imposed either
$p=0$ or $q=\text{const}$ on the boundary as BC for on-shell fields,
in other words as the BC which make the field equations to a
mathematically well-posed problem. Should one also use these BC
for off-shell fields, for example in path integrals? We do not
believe so as it is natural to preserve ``susy without BC.'' 
If one does impose BC off-shell, the boundary action can be
simplified (and in our examples it would vanish), but the resulting
bulk-plus-boundary action would not be ``susy without BC.''
If the boundary terms in the EL variation of the action are not
of the form ``$p \da q$'' (but in the cases we studied
they could always be cast into this form by adding a suitable
separately susy action on the boundary),
we believe that any set of BC, which
makes this boundary term vanish \emph{on-shell}, is allowed.
Taking any set of (on-shell) BC requires, of course, to study
their consistency and to construct the orbit of BC.
This orbit is particularly simple in our ``susy without BC''
formulation: then the orbit is just a boundary superfield 
(provided we keep enough auxiliary fields).

The results of the present article for rigidly susy models
in $x$-space and superspace, and those of \cite{dbpvn1} for
locally susy models in $x$-space, have settled some of the questions
we had about susy models with boundaries.
We are now interested in tackling the Horava-Witten model
in 11D \cite{hw,moss}, and various (susy) AdS/CFT and Randall-Sundrum
models in dimensions greater than four.
In these cases full sets of auxiliary fields are not known
(or do not exist), and thus no complete superspace formulations
are available. Therefore, many of our constructions are not directly
applicable. However, in our articles we also studied the issue
of eliminating auxiliary fields while preserving ``susy without BC,''
and in many cases it was indeed possible to do so. Therefore, we
expect that some of the higher dimensional models can be made
``susy without BC.''

\vspace{30pt}
{\bf Acknowledgments.}
We thank the C. N. Yang Institute
for Theoretical Physics at SUNY Stony Brook and Deutsches
Electronen-Synchrotron DESY in Hamburg for hospitality extended
to us during visits related to this project. The research of D.V.B.
was supported in part by the German Science Foundation (DFG).
The research of P.v.N. was supported by the NSF grant no. PHY-0354776.



\end{document}